\documentstyle[amsmath,amssymb,graphicx]{article}

\def\be{\begin{eqnarray}}
\def\ee{\end{eqnarray}}
\def\nn{\nonumber}

\def\T{\hbox{{\bf T}}}
\def\G{{\bf\hbox{G}}}

\def\p{\partial}

\def\Tr{{\rm Tr}\,}

\def\spepo{\sigma}

\hoffset= - 0.9in          

\title{{\bf Equations on knot polynomials
and 3d/5d  duality } \vspace{.2cm}}
\author{{\bf A.Mironov}\footnote{ {\small {\it
Lebedev Physics Institute} and {\it ITEP, Moscow, Russia}};
mironov@itep.ru; mironov@lpi.ru}\hspace{2mm} and {\bf A.Morozov}\thanks{{\small
{\it ITEP, Moscow, Russia}}; morozov@itep.ru}\date{ }}

\begin{document}
 \maketitle

\vspace{-5cm}

\begin{center}
\hfill FIAN/TD-17/12\\
\hfill ITEP/TH-34/12\\
\end{center}

\vspace{3.5cm}

\centerline{ABSTRACT}

\bigskip

{\footnotesize We briefly review the current situation with various relations
between knot/braid polynomials (Chern-Simons correlation functions),
ordinary and extended, considered as functions of the representation
and of the knot topology.
These include linear skein relations, quadratic Plucker relations,
as well as "differential" and (quantum) $\hat{\cal A}$-polynomial structures.
We pay a special attention to identity between the $\hat{\cal A}$-polynomial equations
for knots and Baxter equations for quantum relativistic integrable systems,
related through Seiberg-Witten theory to $5d$ super-Yang-Mills  models
and through the AGT relation to the $q$-Virasoro algebra.
This identity is an important ingredient of emerging a $3d-5d$ generalization
of the AGT relation.
The shape of the Baxter equation (including the values of coefficients)
depend on the choice of the knot/braid.
Thus, like the case of KP integrability, where (some, so far torus) knots
parameterize particular points of the Universal Grassmannian,
in this relation they parameterize particular points in the
moduli space of many-body integrable systems of relativistic type.}

\vspace{.5cm}

\section{Ordinary and extended HOMFLY polynomials}

The HOMFLY polynomials \cite{HOMFLY}
\be
\left.H_R^{\cal K}(q|A)\right|_{A=q^N}\ = \
\Big<\Tr_R P\exp\oint_{\cal K}{\mathcal A}\Big>_{SU(N)}
\ee
are the analytical continuation in $N$ of the Wilson-loop
averages in $3d$ Chern-Simons theory \cite{CS} with the gauge group $SU(N)$
and the action
$S = {\kappa\over 4\pi} \int_{d^3x} \Tr\Big({\mathcal A}d{\mathcal A}
+ \frac{2}{3}{\mathcal A}^3\Big)$,
where $q = \exp\frac{2\pi i}{\kappa+N}$.
They can be lifted to {\it extended} polynomials \cite{I,II}:
${\cal H}_R^{\cal B}\{q|p_k\}$, which depend on
infinitely many time-variables $\{p_k\}$
and on particular braid representation ${\cal B}$
of the oriented knot ${\cal K}$
(the braid arises in projection onto a $2d$ plane and depends on the
particular embedding of the knot and on the choice of the plane),
so that
\be
H_R^{\cal K}(q|A) = \Big(q^{-4\varkappa_R} A^{-|R|}\Big)^{\#({\cal B})}
{\cal H}_R^{\cal B}\{q|p_k^*\}
\label{HvsextH}
\ee
where $\#({\cal B})$ is the writhe number, i.e. the algebraic number
of crossings in the braid ${\cal B}$, and
\be
p_k^* = \frac{A^k-A^{-k}}{q^k-q^{-k}}
\label{tolo}
\ee
At this {\it topological locus} the r.h.s. of (\ref{HvsextH})
does not depend on the choice of ${\cal B}$ for a given ${\cal K}$.

An $m$-strand braid is parameterized by a sequence of integers $a_{ij}$
with $i=1,2,\ldots,m-1$ and $j=1,\ldots,n$, their meaning can be understood from the picture
(in this figure $a_1=-2$, $b_1=2$, $a_2=-1$, $b_2=3$:
this is knot $8_{10}$):

\vspace{0.8cm}

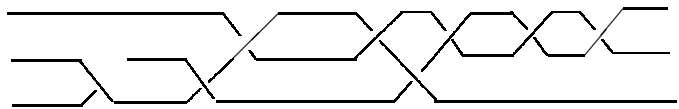
\unitlength 0.7mm 
\linethickness{0.6pt}
\ifx\plotpoint\undefined\newsavebox{\plotpoint}\fi 
\begin{picture}(45.5,53)(-40,-10)
\put(19.5,34.5){\line(1,0){13.25}}
\put(41.25,43.25){\line(1,0){11.25}}
\put(19.25,43){\line(1,0){13.25}}
\put(38.75,35){\line(1,0){13.75}}
\put(61.25,43.25){\line(1,1){8.75}}
\put(70,52){\line(1,0){14.75}}
\put(18.5,52){\line(1,0){41}}
\multiput(59.5,52)(.033505155,-.043814433){97}{\line(0,-1){.043814433}}
\put(58.25,35.25){\line(1,0){33.75}}
\multiput(92,35.25)(.033505155,.038659794){97}{\line(0,1){.038659794}}
\multiput(64.5,45)(.03289474,-.04605263){38}{\line(0,-1){.04605263}}
\put(65.75,43.25){\line(1,0){19}}
\multiput(84.5,43.5)(.0346153846,.0336538462){260}{\line(1,0){.0346153846}}
\multiput(84.75,52)(.03370787,-.03651685){89}{\line(0,-1){.03651685}}
\multiput(52.5,43)(.033653846,-.046474359){156}{\line(0,-1){.046474359}}
\multiput(52.5,35)(.03353659,.03353659){82}{\line(0,1){.03353659}}
\multiput(56.75,39)(.035447761,.03358209){134}{\line(1,0){.035447761}}
\multiput(32.25,43)(.033602151,-.041666667){186}{\line(0,-1){.041666667}}
\multiput(32.75,34.75)(.03333333,.03333333){75}{\line(0,1){.03333333}}
\put(37,39){\line(1,1){4.25}}
\put(99.75,35.25){\line(1,0){45.75}}
\multiput(100,35.5)(-.0336990596,.0352664577){319}{\line(0,1){.0352664577}}
\multiput(97.25,41)(.0336363636,.04){275}{\line(0,1){.04}}
\put(106.5,52){\line(1,0){7.75}}
\put(121.25,44){\line(1,0){6.75}}
\put(128,44){\line(5,6){7.5}}
\put(135.5,53){\line(1,0){8.25}}
\put(93.25,52.25){\line(1,0){5.75}}
\multiput(99,52.25)(.03353659,-.04268293){82}{\line(0,-1){.04268293}}
\multiput(103,47)(.03333333,-.05){60}{\line(0,-1){.05}}
\put(105,44){\line(0,1){0}}
\put(105,44){\line(1,0){9.5}}
\multiput(114.5,44)(.033632287,.036995516){223}{\line(0,1){.036995516}}
\put(122,52.25){\line(1,0){5.25}}
\multiput(127.25,52.25)(.03353659,-.03963415){82}{\line(0,-1){.03963415}}
\multiput(131.5,47)(.03333333,-.04166667){60}{\line(0,-1){.04166667}}
\put(133.5,44.5){\line(1,0){10.75}}
\multiput(114.25,52.25)(.03370787,-.03651685){89}{\line(0,-1){.03651685}}
\multiput(121,44)(-.03333333,.04666667){75}{\line(0,1){.04666667}}
\end{picture}

\vspace{-2.7cm}

Representation is parameterized by the Young diagram: an ordered set
of $l(R)$ positive integers $R = \{r_1\geq r_2\geq \ldots \geq r_{l(R)}>0\}
= [r_1,r_2,\ldots,r_{l(R)}]$.

One of the hot questions in Chern-Simons theory is to find and investigate
the equations, which the HOMFLY and extended HOMFLY polynomials satisfy as
functions of $a_{ij}$ and $r_k$.

The main source of information about these relations
(besides pure empiricism) is the {\it character expansion}
separating dependencies on the braid and the time variables\footnote{This presentation
is essentially based on earlier attempts in \cite{TR,chi,indians}.}:
\be
{\cal H}^{\cal B}_R\{q|p_k\} = \sum_{Q\vdash m|R|} c_{RQ}^{\cal B}(q) S_Q\{p_k\}
\label{Hexpan}
\ee
where $S_Q$ are the Schur functions, i.e. the characters of $GL(\infty)$ in representations
(Young diagrams) $Q$ of the size $|Q|=m|R|$ and $c_{RQ}^{\cal B}(q)$
are represented as traces in the spaces of intertwining operators
of combinations of simple matrices,
\be
c_{RQ}^{\cal B}(q) = \Tr_Q \prod_j
\left(\prod_{i=1}^{m-1}\hat{\cal R}^{a_{ij}}\hat{\cal U}_i\right)
\label{casTr}
\ee
described in detail in refs.\cite{I,II,III,Ano}.
More artful character expansions (especially in the Hall-Littlewood or MacDonald polynomials)
can be even more useful for particular applications \cite{DMMSS,MMSS,MMS}.

In this paper we briefly review the main known
linear and bilinear relations between the HOMFLY polynomials,
as well as some closely related topics.
The subject is still at empirical stage of development,
far less understood than, say, Virasoro constraints, various recursions
and integrability in matrix models, though a list of known properties and facts
is quite impressive. Therefore, it deserve an attention and further deep investigation.

\section{Skein relations for ordinary and colored HOMFLY polynomials}

The oldest known equation of this type is the {\it skein relation},
which for {\it extended} polynomials states that
\be
{\cal H}_{_\Box}^{(\ldots\ a_{ij}+1\ \ldots)} -
{\cal H}_{_\Box}^{(\ldots\ a_{ij}-1\ \ldots)}
= \left(q-q^{-1}\right) {\cal H}_{_\Box}^{(\ldots\ a_{ij}\ \ldots)}
\label{skefun}
\ee
for any particular parameter $a_{ij}$.
The origin of this relation is that the ${\cal R}$-matrix, acting in
the channel $\underline{[1]}\otimes\underline{[1]} = \underline{[2]}\oplus\underline{[11]}$ has two eigenvalues,
$q$ and $-1/q$, associated with the two irreducible representations
$\underline{[2]}$ and $\underline{[11]}$ respectively, and thus satisfies the Hecke algebra
constraint
\be\label{7}
\Big({\cal R} - q\Big)\Big({\cal R}+q^{-1}\Big) = 0
\ee
Eq.(\ref{skefun}) is then a direct corollary of representation (\ref{casTr}).

For bigger representations $R$, the product
$R\otimes R = \oplus_S S$ and ${\cal R}$-matrix in this channel
has many different eigenvalues: as many as there are irreducible
representations $S$ in this expansion.
These eigenvalues are equal to $\epsilon_Sq^{\varkappa_S}$,
where $\varkappa_S$ is an eigenvalue of the cut-and-join operator \cite{cj}:
$\hat W_{[2]}$, associated with its eigenfunction character $S_S\{p_k\}$,
and $\epsilon_S=\pm 1$, depending on $R$ and $S$.
Therefore, this ${\cal R}$-matrix satisfies
\be\label{8}
\prod_S \Big({\cal R} - \epsilon_Sq^{\varkappa_S}\Big) = 0
\ee
and, as a corollary of (\ref{casTr}),
\be\label{9}
\prod_S \left(e^{\p/\p a_{ij}} - \epsilon_Sq^{\varkappa_S}\right)
{\cal H}_R^{(\ldots a_{ij} \ldots)} = 0
\ee
for any variable $a_{ij}$.

For example, in the case of the first symmetric representation, one has
\be
\Big({\cal R} - q^6\Big)\Big({\cal R}+q^{2}\Big)\Big({\cal R} - 1\Big) = 0
\ee
i.e.
\be
{\cal H}_{[2]}^{(\ldots\ a_{ij}+3\ \ldots)} - (1+q^2+q^6)
{\cal H}_{[2]}^{(\ldots\ a_{ij}+2\ \ldots)} + q^2(1+q^4+q^6)
{\cal H}_{[2]}^{(\ldots\ a_{ij}+1\ \ldots)}
- q^8{\cal H}_{[2]}^{(\ldots\ a_{ij}\ \ldots)}  = 0
\ee
while for the antisymmetric representation
\be
{\cal H}_{[11]}^{(\ldots\ a_{ij}+3\ \ldots)} - (1+q^{-2}+q^{-6})
{\cal H}_{[11]}^{(\ldots\ a_{ij}+2\ \ldots)} + q^{-2}(1+q^{-4}+q^{-6})
{\cal H}_{[11]}^{(\ldots\ a_{ij}+1\ \ldots)}
- q^{-8}{\cal H}_{[11]}^{(\ldots\ a_{ij}\ \ldots)}  = 0
\ee

\bigskip

To make use of the skein relations for evaluation of the HOMFLY polynomials
one needs to supplement them by "the initial conditions":
in the case of $R=\Box$ one can express arbitrary ${\cal H}^{a_{ij}}$
with arbitrary $\{a_{ij}\}$ through those where all $a_{ij}=0$ or $1$.
These polynomials should be found by some other means (e.g., at topological
locus topological invariance can be used to decrease the number of strands).
For higher representations the "boundary" $a_{ij}$ can take more values.

For superpolynomials \cite{sp,DGR} the skein relations are non-trivially deformed.
Already for the simplest $2$-strand torus knots and links only one
half of the relations remain intact:
\be
{\cal P}^{[2,n+1]}_{_\Box} - {\cal P}^{[2,n-1]}_{_\Box}=
\Big(q-q^{-1}\Big) {\cal P}^{[2,n]}_{_\Box}
\ee
for odd $n$ (so that there are links at one of the sides of the relation),
but for even $n$ this is not true.
The origin of this phenomenon is that the character expansion
(in MacDonald polynomials) is different for knots and links \cite{DMMSS}:
\be
{\cal P}_{_\Box}^{[2,n]} =
q^n M^*_{[2]}\{p\} \ \ + \ \
\left\{\begin{array}{cccccc} -&  \frac{(1+q^2)(1-t^2)}{1-q^2t^2}\cdot t^{-n} M_{[11]}\{p\}
& {\rm for} & n & {\rm odd} & ({\rm knots}) \\ \\
+ & \frac{t(1-q^4)}{q(1-q^2t^2)}\cdot t^{-n}M_{[11]}\{p\}
& {\rm for} & n & {\rm even} & ({\rm links})
\end{array} \right.
\ee

\section{Other relations for knot-dependence}

Skein relations restrict the dependence of the knot polynomial on the shape of the braid.
They are the only {\it universal} equations of this kind, known so far.
However, it is already clear that they are not the only ones:
some traces of a more profound structure are already observed.
We give just two examples, both for torus knots and links.

\subsection{"Evolution" of knots induced by skein relations}

\subsubsection*{Torus knots/links}

Skein relation as it is can not be directly deals with only torus knots,
except for the $2$-strand case: for the number of strands $m>2$
it mixes torus knots with non-torus ones.
However, the simple procedure, described in the previous section,
immediately provides another kind of relations which provides
a set of skein relations closed on only the $m$-strand torus knots.
Namely, the HOMFLY polynomial for the $m$-strand torus knot/link $[m,n]$
is a weighted trace of the $n$-the power of a special combination
of $m-1$ ${\cal R}$-matrices,
\be
{\cal H}^{[m,n]}_R\{p\} = \Tr \check{\cal R}_m^n,
\ \ \ \
\check{\cal R}_m = {\cal R}_{12}{\cal R}_{23}\ldots {\cal R}_{m-1,m}
\ee
If instead of an individual ${\cal R}$ in the standard
skein relation, one now diagonalizes $\check{\cal R}_m$,
one can immediately write the analogues of eqs.(\ref{8}) and (\ref{9})
for the series of knots $[m,n]$ with given $m$.
For example, for the $3$-strand torus links in the fundamental
representation one has \cite{II}
\be
{\cal H}^{[3,n]}_{\Box}\{p\} = q^{2n} S_{[3]}\{p\}
+ \Tr_{2\times 2}\Big(\hat{\cal R}\hat U\hat{\cal R}\hat U^\dagger\Big)^n
\cdot S_{[21]}\{p\} + (-1/q)^{2n}S_{[111]}\{p\}
\ee
and the $2\times 2$ matrix $\hat{\cal R}\hat U\hat{\cal R}\hat U^\dagger$
has eigenvalues $e^{\pm 2\pi i/3}$.
This means that our ${\cal R}_3$ in this case has four different
eigenvalues: $q^2, e^{\pm 2\pi i/3}, q^{-2}$, i.e. satisfies
\be\label{16}
\Big({\cal R}_3-q^2\Big)\Big({\cal R}_3-q^{-2}\Big)
\Big({\cal R}_3^2+{\cal R}_3+1\Big) = 0
\ee
and this implies the equation for torus HOMFLY polynomial:
\be
\Big(e^{\partial/\partial n}-q^2\Big)\Big(e^{\partial/\partial n}-q^{-2}\Big)
\Big(e^{2\partial/\partial n}+e^{\partial/\partial n}+1\Big){\cal H}^{[3,n]}_{\Box}\{p\}
= \nn \\ =
\Big({\cal H}^{[3,n+4]}_{\Box}\{p\} + {\cal H}^{[3,n+3]}_{\Box}\{p\}
+ {\cal H}^{[3,n+2]}_{\Box}\{p\}\Big)-\nn\\
- (q^2+q^{-2})\Big({\cal H}^{[3,n+3]}_{\Box}\{p\} + {\cal H}^{[3,n+2]}_{\Box}\{p\}
+ {\cal H}^{[3,n+1]}_{\Box}\{p\}\Big)
+ \nn \\
+ \Big({\cal H}^{[3,n+2]}_{\Box}\{p\} + {\cal H}^{[3,n+1]}_{\Box}\{p\}
+ {\cal H}^{[3,n]}_{\Box}\{p\}\Big) = 0
\ee
Thus one obtains a kind of evolution in the number of $\check{\cal R}_m$. This evolution can be described
by "an evolution operator" \cite{DMMSS}, which is nothing but the cut-and join operator \cite{cj}. Moreover, this
evolution can be naturally continued to the superpolynomials \cite{DMMSS,MMSS}.

If one wants instead an equation for the topological invariant
HOMFLY polynomial, it is necessary to restrict time-variables to topological
locus (\ref{tolo}), $p_k=p_k^*$ and also include the proper normalization factor (\ref{HvsextH}). In this case
the evolution can be considered as an evolution in the number of strands.

\subsubsection*{Non-torus knots/links}

In fact, in just the same way one can write a version of skein relation
for an arbitrary one-parametric family of braids, obtained by iteration
of any braid group element: ${\cal B}^{(n)} = {\cal B}_0{\cal B}_1^n$.
If the product ${\cal R}_1$ of ${\cal R}$-matrices, associated with ${\cal B}_1$,
has eigenvalues $\xi_i$, then
\be
\prod_i \Big({\cal R}_1 - \xi_i(q)\Big) = 0
\ee
and the corresponding extended HOMFLY polynomials
satisfy
\be
\prod_i \Big(e^{\partial/\partial n} - \xi_i(q)\Big){\cal H}^{{\cal B}^{(n)}}\{p\} = 0
\ee
independently of the choice of the "initial" braid element ${\cal B}_0$. Thus, one can start an evolution with any
"initial condition" ${\cal B}_0$.
For examples of such sequences see \cite{DMMSS,II}, note that
eigenvalues can be somewhat non-trivial functions of $q$.

Generalization of this procedure to superpolynomials seems straightforward,
but it is still unclear if it really works in full generality,
see \cite{DMMSS}.

\subsection{Strand-changing relations}

Like the skein relation itself, the modification, considered in the previous
subsection does not change the number of strands in the braid.
However, the strand-changing is also possible.
An example is the following relation, which at the moment is very restricted:
only torus knots, only $SU(2)$ gauge group $(A=q^2$, i.e. Jones polynomials),
only fundamental representation, but instead
it is in a very different direction in the braid-parameter space,
and it survives (with minor modifications) a $\beta$-deformation
to superpolynomials \cite{ST}.

Namely, for torus knots of the special type $[m,m+1]$
($m$-strand braids, but a specially restricted $n=m+1$)
\be
{\cal H}^{[m,m+1]}_{_\Box}\{p\} = q^{m^2-1}S_{[m]}\{p\} -
q^{(m-3)(m+1)}S_{[m-1,1]}\{p\} + \ldots
\ee
where the sum goes over all hook diagrams of the size $m$.
However, on the Jones topological locus,
\be
p_k = p_k^* = \frac{A^k-A^{-k}}{q^k-q^{-k}}
\ \stackrel{A=q^2}{\longrightarrow} q^k+q^{-k}
\ee
only
$ S_{[n]}^* = \frac{\{q^{n+1}\}}{\{q\}}$
and
$
S_{[n-1,1]}^* = S_{[n-2]}^* = \frac{ \{ q^{n-1} \}} {\{q\}}
$
are non-vanishing so that
\be
H^{[m,m+1]}_{_\Box}(A=q^2|q) =
A^{-(m^2-1)}
{\cal H}^{[m,m+1]}_{_\Box}\{p^*\} =
{q^{-(m+1)(m+2)}\over q-q^{-1}}
\ \Big(q^{4m+4}-q^{2m+2}-q^{2m} + q^2\Big)\equiv\nn\\ \equiv {q^{-(m+1)(m+2)}\over q-q^{-1}}{\cal J}^{(m)}
\ee
The four-term polynomial ${\cal J}^{(m)}(q)$ satisfies the
$4$-term recurrent relation
\be
{\cal J}^{(m)} = q^4 {\cal J}^{(m-1)} + q^{-2m+8}\Big({\cal J}^{(m-2)} - {\cal J}^{(m-3)}\Big)
\ee
For generalizations to Jones unreduced superpolynomials for the same series
of knots see \cite{GOR} (where slightly different normalization is used and $q\to 1/q$;
hereafter, such a difference with expressions for the trefoil in the literature arises, because we use the right-hand trefoil
described by braid $(1,1|1,1)$, while often the left-hand trefoil $(-1,-1|-1,-1)$ is used instead).

In fact, there is nothing special about the $[m,m+1]$ series,
at least, for the HOMFLY polynomials.
For $[m,m+r]$ one similarly gets
\be
H^{[m,m+r]}_{_\Box}(A=q^2|q) = q^{-2(m-1)(m+r)}
\Big( q^{(m-1)(m+r)}S_{[m]}^* - q^{(m-3)(m+r)}S_{[m-2]}^*\Big)
= \frac{q^{-(m+1)(m+1+r)}}{q-q^{-1}}
{\cal J}^{(m|r)}
\ee
where the 4-term (differently normalized) Jones
polynomial
\be
{\cal J}^{(m|r)} = q^{4m+2r+2} - q^{2m+2r} - q^{2m} +q^2
\ee
satisfies
\be
{\cal J}^{(m|r)} - q^{4r}{\cal J}^{(m-r|r)} =
q^{-2m+2r+6}\Big({\cal J}^{(m-2|r)} - {\cal J}^{(m-2-r|r)}\Big)
\ee
Note that this relation at some values of $m$ and $r$ involves both the Jones polynomials of
knots and of links. For instance, at $m=7$ and $r=3$, $T[7,10]$, $T[4,7]$, $T[5,8]$ and $T[2,5]$
are involved, all of them being knots, while at $m=8$ and $r=3$, of involved $T[8,11]$, $T[5,8]$, $T[6,9]$
and $T[3,6]$ only the first two are knots, while the second two are links.

\subsection{Colored HOMFLY by cabling}

Another application of (\ref{7}) and (\ref{16}) is to construct colored HOMFLY polynomials from the ordinary
ones. To this end, as a first step, one has to use the cabling procedure, i.e. the relation
\be
H_{R^{\otimes m}}^{{\cal K}}=H_R^{{\cal K}^m}
\ee
where ${\cal K}^m$ denotes $m$-cabling of the knot ${\cal K}$
(see, e.g., \cite[eq.(1.5b)]{DG}). At the same time, the l.h.s. of this formula can be presented as a sum over the
irreducible representations,
\be
H_{R^{\otimes m}}^{{\cal K}}=\sum_SH_{S}^{{\cal K}}
\ee
Therefore, in order to calculate $H_{S}^{{\cal K}}$, one has to calculate ${\cal K}^m$ in the fundamental representation
and then project out the result onto an irreducible representation $Q$. This is done using proper projectors.

For instance, relation (\ref{7}) implies that
\be
P_{[2]}={1+q{\cal
R}\over 1+q^2}\ \ \ \ \ \ \hbox{and}\ \ \ \ \ \ P_{[11]}={q^2-q{\cal
R}\over 1+q^2}
\ee
are projectors correspondingly onto
representations $[2]$ and $[11]$, which emerge in the decomposition
of the product $[1]\times [1]$. Similarly, (\ref{16}) implies that
the projectors onto irreducible components of the triple product
$[1]\times [1]\times [1]=[3]+2[21]+[111]$ are
\be
P_{[3]}={(1-q^2{\cal R}_3)({\cal R}^2_3+{\cal R}_3+1)\over (1-q^4)(1+q^2+q^4)}\
,\ \ \ \ \ \ \ P_{[111]}=q^6{({\cal R}-q^2)({\cal R}^2_3+{\cal
R}_3+1)\over (1-q^4)(1+q^2+q^4)}
\ee
and
\be
P_{[21]}^{\pm}=\pm{e^{\pm \pi i/6}\over\sqrt{3}}
{(q^2{\cal R}_3-1)({\cal R}_3-q^2)({\cal R}_3-e^{\pm 2\pi i/3})\over 1+q^2+q^4}
\ee
projects on two representations $[21]$ that correspond to the two eigenvalues $e^{\pm 2\pi i/3}$ of ${\cal R}_3$.

\section{Differentials}

Another type of relations comes from the trivial observation that
\be
H_R^{\cal K}(A|q) - H_R^{\cal K}(q^N|q) = (A-q^N)G_R^{\cal K}(A|q),\nn\\
H_R^{\cal K}(A|q) - H_R^{\cal K}(-(1/q)^N|q) = \Big(A-(-1/q)^N\Big)\tilde G_R^{\cal K}(A|q)
\ee
with some polynomials $G_R^{\cal K}(A|q)$, $\tilde G_R^{\cal K}(A|q)$.
This naturally leads to the formalism of "differentials", widely used in application
to {\it superpolynomials}.
It is effective, when an additional information is used about the HOMFLY polynomials
for a particular $N$, for example, that $S_Q^* = S_Q\{p_k^*\}$ in (\ref{Hexpan})
are non-vanishing only for $l(Q)\leq N$ (in particular, for $N=1$ a single
term with $Q=[m|R|]$ survives in the character expansion of
$H_R^{\cal K}(q|q)$).

These relations have a nice pictorial representation \cite{DGR}.
Let us plot the HOMFLY polynomials $H_R^{\cal K}(A|q)$ as a set of points corresponding to
the monomials $A^kq^l$
on the $(k,l)$-plane with positive and negative signs
denoted by the white and black circles respectively, and with multiplicities (if necessary).
Then, all the terms that could cancel under the specialization $A=q^N$,
lie on the lines $Nk=-l+const$, and they cancel if the algebraic sum of the terms on each line vanish.
Similarly, all the terms that cancel under the specialization $A=-1/q^N$
lie on the lines $Nk=l+const$, and the algebraic sum of the terms on each such line should
also vanish.
An important point is that this line structure is preserved
by a $\beta$-deformation, and this provides an important
tool to deal with the superpolynomials \cite{DGR}.

As the simplest example, the fundamental HOMFLY polynomials for all the 2-strand knots
pictorially look basically the same: since
\be
\frac{H_{\Box}^{[2,2n+1]}(A|q)}{S_{_\Box}^*}
\sim \frac{q^{2n+2}-q^{-2n-2}}{q^2-q^{-2}}A
- \frac{q^{2n}-q^{-2n}}{q^2-q^{-2}}A^{-1}
\ee
they are sets of $n+1$ and $n$ points on the horizontal lines $k=+1$ and $k=-1$ respectively, and
one has, say, for the $[2,9]$ case, for
$(q^8+q^4+1+q^{-4}+q^{-8})A - (q^6+q^2+q^{-2}+q^{-6})A^{-1}$

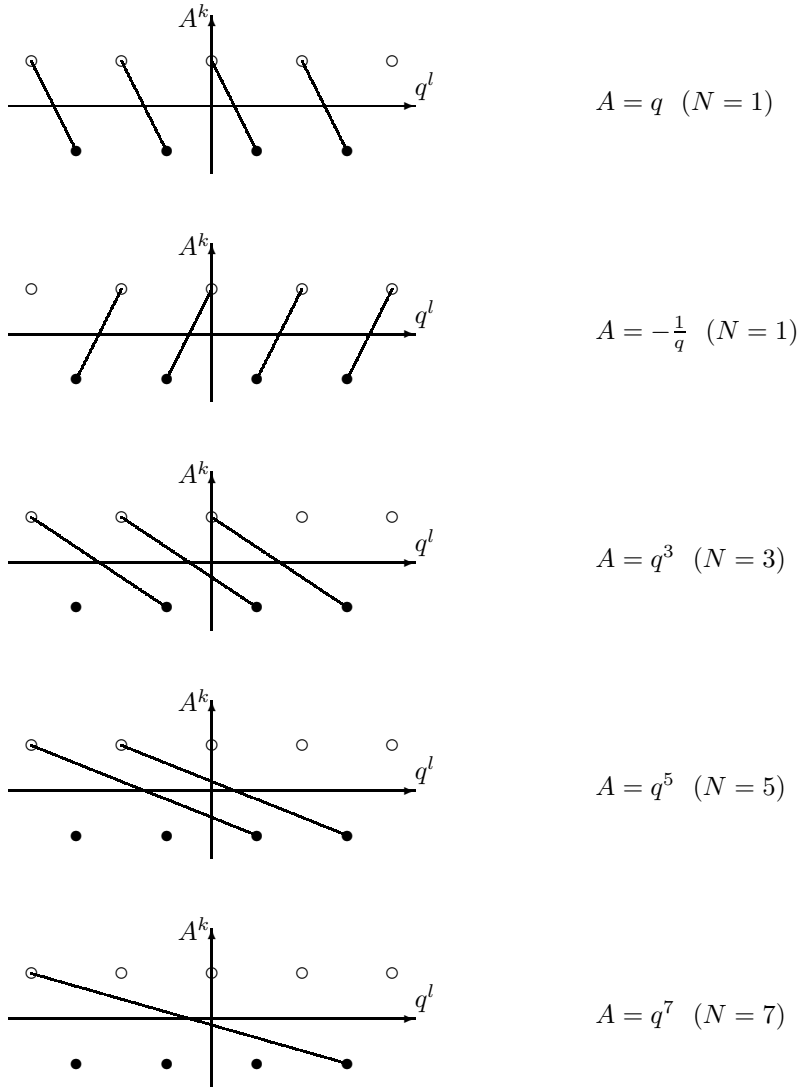
\unitlength 0.3mm 
\begin{picture}(300,100)(-70,-40)
\put(0,20){\circle{5}}
\put(40,20){\circle{5}}
\put(80,20){\circle{5}}
\put(120,20){\circle{5}}
\put(160,20){\circle{5}}
\put(20,-20){\circle*{5}}
\put(60,-20){\circle*{5}}
\put(100,-20){\circle*{5}}
\put(140,-20){\circle*{5}}
\put(-10,0){\vector(1,0){180}}
\put(80,-30){\vector(0,1){70}}
\put(65,35){\mbox{$A^k$}}
\put(170,5){\mbox{$q^l$}}
\qbezier(20,-20)(10,0)(0,20)
\qbezier(60,-20)(50,0)(40,20)
\qbezier(100,-20)(90,0)(80,20)
\qbezier(140,-20)(130,0)(120,20)
\put(250,-2){\mbox{$A=q\ \ (N=1)$}}
%
%
%
%
%
\end{picture}

\begin{picture}(300,100)(-70,-40)
\put(0,20){\circle{5}}
\put(40,20){\circle{5}}
\put(80,20){\circle{5}}
\put(120,20){\circle{5}}
\put(160,20){\circle{5}}
\put(20,-20){\circle*{5}}
\put(60,-20){\circle*{5}}
\put(100,-20){\circle*{5}}
\put(140,-20){\circle*{5}}
\put(-10,0){\vector(1,0){180}}
\put(80,-30){\vector(0,1){70}}
\put(65,35){\mbox{$A^k$}}
\put(170,5){\mbox{$q^l$}}
\qbezier(20,-20)(30,0)(40,20)
\qbezier(60,-20)(70,0)(80,20)
\qbezier(100,-20)(110,0)(120,20)
\qbezier(140,-20)(150,0)(160,20)
\put(250,-2){\mbox{$A=-\frac{1}{q}\ \ (N=1)$}}
\end{picture}

\begin{picture}(300,100)(-70,-40)
\put(0,20){\circle{5}}
\put(40,20){\circle{5}}
\put(80,20){\circle{5}}
\put(120,20){\circle{5}}
\put(160,20){\circle{5}}
\put(20,-20){\circle*{5}}
\put(60,-20){\circle*{5}}
\put(100,-20){\circle*{5}}
\put(140,-20){\circle*{5}}
\put(-10,0){\vector(1,0){180}}
\put(80,-30){\vector(0,1){70}}
\put(65,35){\mbox{$A^k$}}
\put(170,5){\mbox{$q^l$}}
\qbezier(60,-20)(30,0)(0,20)
\qbezier(100,-20)(70,0)(40,20)
\qbezier(140,-20)(110,0)(80,20)
%
\put(250,-2){\mbox{$A=q^3\ \ (N=3)$}}
\end{picture}

\begin{picture}(300,100)(-70,-40)
\put(0,20){\circle{5}}
\put(40,20){\circle{5}}
\put(80,20){\circle{5}}
\put(120,20){\circle{5}}
\put(160,20){\circle{5}}
\put(20,-20){\circle*{5}}
\put(60,-20){\circle*{5}}
\put(100,-20){\circle*{5}}
\put(140,-20){\circle*{5}}
\put(-10,0){\vector(1,0){180}}
\put(80,-30){\vector(0,1){70}}
\put(65,35){\mbox{$A^k$}}
\put(170,5){\mbox{$q^l$}}
\qbezier(100,-20)(50,0)(0,20)
\qbezier(140,-20)(90,0)(40,20)
%
\put(250,-2){\mbox{$A=q^5\ \ (N=5)$}}
\end{picture}

\begin{picture}(300,100)(-70,-40)
\put(0,20){\circle{5}}
\put(40,20){\circle{5}}
\put(80,20){\circle{5}}
\put(120,20){\circle{5}}
\put(160,20){\circle{5}}
\put(20,-20){\circle*{5}}
\put(60,-20){\circle*{5}}
\put(100,-20){\circle*{5}}
\put(140,-20){\circle*{5}}
\put(-10,0){\vector(1,0){180}}
\put(80,-30){\vector(0,1){70}}
\put(65,35){\mbox{$A^k$}}
\put(170,5){\mbox{$q^l$}}
\qbezier(140,-20)(70,0)(0,20)
%
\put(250,-2){\mbox{$A=q^7\ \ (N=7)$}}
\end{picture}

\noindent
The circles not connected by oblique lines survive to describe
the specialized polynomials $H_{_\Box}(A=q^N|q)$.
For even $N$ and for odd $N\geq 9$
no reduction takes place: all the nine items are present.

Similarly, for the figure eight knot $4_1$ with
$H_{_\Box}^{(1,-1|1,-1)} = 1 + \{Aq\}\{A/q\} = A^2 -q^2+1-q^{-2} + A^{-2}$ one has

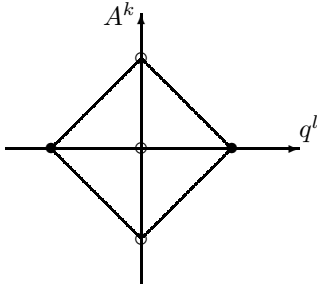
\begin{picture}(300,150)(-90,-70)
\put(0,-40){\circle{5}}
\put(0,0){\circle{5}}
\put(0,40){\circle{5}}
\put(-40,0){\circle*{5}}
\put(40,0){\circle*{5}}
%
\put(-60,0){\vector(1,0){130}}
\put(0,-60){\vector(0,1){120}}
\put(-17,55){\mbox{$A^k$}}
\put(70,5){\mbox{$q^l$}}
\qbezier(0,-40)(-20,-20)(-40,0)
\qbezier(0,-40)(20,-20)(40,0)
\qbezier(0,40)(-20,20)(-40,0)
\qbezier(0,40)(20,20)(40,0)
\put(180,-2){\mbox{$\begin{array}{c}A=q\\A=-1/q\end{array}\ \ (N=1)$}}
\end{picture}

\noindent
and there are no non-trivial reductions, except for $A=q$ and $A=-1/q$ ($N=1$).

\bigskip

One can also consider "unreduced" HOMFLY "polynomials",
not divided by $S_R^*$.
If $A$ is considered as an independent variable, they are actually series,
not polynomials in $q$ and their pictorial representation contains
(semi)infinitely many vertices, of which finitely many survive after specializations
$A=q^N$ with arbitrary $N=1,2,3, \ldots$
For example, already for the trefoil $3_1$ one has
\be
H_{_\Box}^{3_1}(A|q) = \overbrace{\underbrace{-q\frac{A-A^{-1}}{1-q^{2}}}_{S_{_\Box}^*(A|q)}
\underbrace{\Big((q^2+q^{-2})A - A^{-1}\Big)}_{\rm reduced}}^{\rm unreduced}
\ee
and

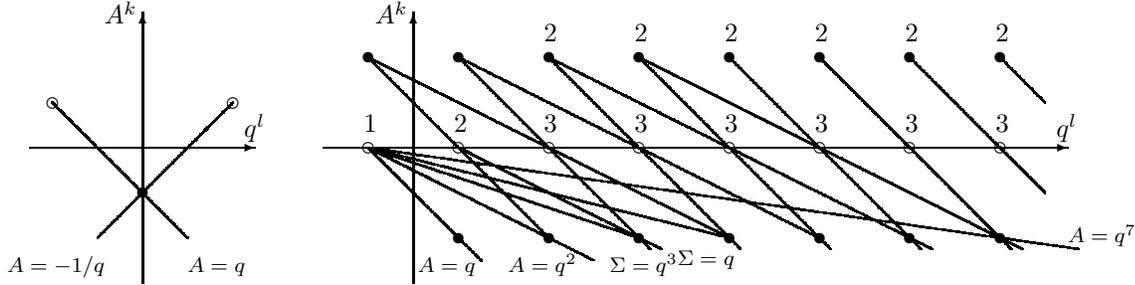
\begin{picture}(300,150)(-160,-70)
\put(-160,20){\circle{5}}
\put(-80,20){\circle{5}}
\put(-120,-20){\circle*{5}}
\put(-170,0){\vector(1,0){100}}
\put(-120,-60){\vector(0,1){120}}
\put(-137,55){\mbox{$A^k$}}
\put(-75,5){\mbox{$q^l$}}
\qbezier(-160,20)(-140,0)(-100,-40)
\qbezier(-80,20)(-100,0)(-140,-40)
\put(-100,-55){\mbox{{\footnotesize $A=q$}}}
\put(-180,-55){\mbox{{\footnotesize $A=-1/q$}}}
\put(20,-40){\circle*{5}}
\put(60,-40){\circle*{5}}
\put(100,-40){\circle*{5}}
\put(140,-40){\circle*{5}}
\put(180,-40){\circle*{5}}
\put(220,-40){\circle*{5}}
\put(260,-40){\circle*{5}}
\put(-20,40){\circle*{5}}
\put(20,40){\circle*{5}}
\put(60,40){\circle*{5}}
\put(100,40){\circle*{5}}
\put(140,40){\circle*{5}}
\put(180,40){\circle*{5}}
\put(220,40){\circle*{5}}
\put(260,40){\circle*{5}}
\put(58,47){\mbox{$2$}}
\put(98,47){\mbox{$2$}}
\put(138,47){\mbox{$2$}}
\put(178,47){\mbox{$2$}}
\put(218,47){\mbox{$2$}}
\put(258,47){\mbox{$2$}}
\put(-20,0){\circle{5}}
\put(20,0){\circle{5}}
\put(60,0){\circle{5}}
\put(100,0){\circle{5}}
\put(140,0){\circle{5}}
\put(180,0){\circle{5}}
\put(220,0){\circle{5}}
\put(260,0){\circle{5}}
\put(-22,7){\mbox{$1$}}
\put(18,7){\mbox{$2$}}
\put(58,7){\mbox{$3$}}
\put(98,7){\mbox{$3$}}
\put(138,7){\mbox{$3$}}
\put(178,7){\mbox{$3$}}
\put(218,7){\mbox{$3$}}
\put(258,7){\mbox{$3$}}
\put(-40,0){\vector(1,0){330}}
\put(0,-60){\vector(0,1){120}}
\put(-17,55){\mbox{$A^k$}}
\put(285,5){\mbox{$q^l$}}
\qbezier(-20,0)(0,-20)(30,-50)
\put(2,-55){\mbox{{\footnotesize $A=q$}}}
\qbezier(-20,40)(20,0)(60,-40)
\qbezier(20,40)(60,0)(105,-45)
\put(87,-55){\mbox{{\footnotesize $\Sigma=q^3$}}}
\qbezier(60,40)(100,0)(145,-45)
\qbezier(100,40)(140,0)(185,-45)
\qbezier(140,40)(180,0)(225,-45)
\qbezier(180,40)(220,0)(265,-45)
\qbezier(220,40)(260,0)(280,-20)
\qbezier(260,40)(270,30)(280,20)
\qbezier(-20,0)(20,-20)(80,-50)
\put(42,-55){\mbox{{\footnotesize $A=q^2$}}}
\qbezier(20,0)(60,-20)(110,-45)
\put(117,-52){\mbox{{\footnotesize $\Sigma=q$}}}
\qbezier(-20,40)(60,0)(140,-40)
\qbezier(20,40)(100,0)(180,-40)
\qbezier(60,40)(140,0)(230,-45)
\qbezier(100,40)(180,0)(270,-45)
\qbezier(180,40)(220,0)(260,-40)
\qbezier(220,40)(260,0)(280,-20)
\qbezier(260,40)(270,30)(280,20)
\qbezier(-20,0)(40,-20)(100,-40)
\qbezier(-20,0)(60,-20)(140,-40)
\qbezier(-20,0)(120,-20)(295,-45)
\put(290,-42){\mbox{{\footnotesize $A=q^7$}}}
%
%
\end{picture}

\noindent
Here we manifestly indicated non-zero sums $\Sigma$
along the lines, which, hence, survive after the specialization.

Deformation to superpolynomials dictated by these pictures
by the rules of \cite{DGR} is actually different for the
reduced and unreduced HOMFLY polynomials, what provides two different
families of superpolynomials: the reduced and unreduced ones.
The former are much simpler and have more profound structures
(in particular, possess character decompositions and can be
{\it extended} to arbitrary time variables \cite{DMMSS,I}),
however, the latter seem to be related to somewhat simpler
versions of Khovanov-Rozhansky homological descriptions.
Of course, the two versions are not independent, for the same example
of the trefoil in the fundamental representation one has
\be
H^{3_1}_{_\Box}(q) =\left\{
\begin{array}{lcl}
{\rm reduced:} &&  P^{3_1}_{_\Box}({\bf a},{\bf q},{\bf t}) = 1+{\bf a}^2{\bf t}{\bf q}^{-2}+{\bf a}^2{\bf q}^2{\bf t}^3 \\
\\
{\rm unreduced:}&&  \bar P^{3_1}_{_\Box}({\bf a},{\bf q},{\bf t}) = \displaystyle{{{\bf a}-{\bf a}^{-1}\over
{\bf q}-{\bf q}^{-1}}\ {\bf a}^2{\bf q}^2{\bf t}^3+
\Big({\bf a}{\bf t}+{1\over {\bf a}}\Big)\Big({{\bf a}^2\over{\bf q}^{2}}-1\Big)}
\end{array}
\right.
\ee
and
\be\label{rur}
\bar P^{3_1}_{_\Box}({\bf a},{\bf q},{\bf t}) = \displaystyle{{{\bf a}-{\bf a}^{-1}\over
{\bf q}-{\bf q}^{-1}}\ {\bf q}^2{\bf a}^2{\bf t}^3
+{{\bf a}+{\bf a}^{-1}{\bf t}^{-1}\over
{\bf q}-{\bf q}^{-1}}\ {1-{\bf q}^2{\bf a}^{-2}\over 1+
{\bf q}^2{\bf a}^{-2}{\bf t}^{-1}}
\Big(P^{3_1}_{_\Box}({\bf a},{\bf q},{\bf t})-{\bf q}^2{\bf a}^2{\bf t}^3\Big)}
\ee

One can see that the unreduced superpolynomial $\bar P^{3_1}_{_\Box}({\bf a},{\bf q},{\bf t})$,
in contrast with the reduced one,
contains non-positive coefficients and only after specialization ${\bf a}={\bf q}^N$
becomes a polynomial with positive coefficients. For instance, at ${\bf a}={\bf q}^2$ (Jones)
\be
\bar P^{3_1}_{_\Box}({\bf a},{\bf q},{\bf t})={\bf q}^3{\bf t}+
{1\over{\bf q}}+{\bf q}^7{\bf t}^3+{\bf q}^5{\bf t}^3
\ee
which, indeed, coincides with the unreduced Khovanov homology \cite{katlas}.

A relation similar to (\ref{rur}) persists for other knots.
For instance, for any "thin" knot \cite{DGR}
\be
\bar P^{\cal K}_{_\Box}({\bf a},{\bf q},{\bf t}) = \displaystyle{{{\bf a}-{\bf a}^{-1}\over
{\bf q}-{\bf q}^{-1}}\ \left({{\bf q}\over{\bf a}}
\right)^{S_{\cal K}}+
{{\bf a}+{\bf a}^{-1}{\bf t}^{-1}\over
{\bf q}-{\bf q}^{-1}}\ {1-{\bf q}^2{\bf a}^{-2}\over 1+{\bf q}^2{\bf a}^{-2}{\bf t}^{-1}}
\Big(P^{\cal K}_{_\Box}({\bf a},{\bf q},{\bf t})-\left({{\bf q}\over{\bf a}}
\right)^{S_{\cal K}}\Big)}
\ee
where $S_{\cal K}$ is the signature of the knot (for the torus knot $T[m,n]$, e.g., $S_{\cal K}=(m-1)(n-1)$).
To make formula simpler, we used here slightly different normalization of superpolynomials.

\section{Pl\"ucker relations (KP integrability)}

The third important kind of equations is related to integrability.
For example, for the generating function of the extended HOMFLY polynomials
\be
Z^{\cal B}_R\{p_k|\bar p_k^{(i)}\}=\left\{\begin{array}{ll}
\sum_R {\cal H}^{\cal B}_R\{q|p_k\}S_R\{\bar p_k\}&\hbox{ for knots}\\
\\
\sum_{R_1\ldots R_l} {\cal H}^{\cal B}_{R_1\ldots R_l}\{q|p_k\}\prod_{i=1}^lS_{R_i}\{\bar p_k^{(i)}\}&\hbox{ for links}
\end{array}\right.
\ee
to be a $\tau$-function of the KP hierarchy in the
$p_k/k$-variables \cite{Plukint}, the
coefficients $\xi_Q = \sum_R C_{RQ}^{\cal B}(q)S_R\{q|\bar p_k\}$ should satisfy the infinite set of
quadratic Pl\"ucker relations:
\be
\xi_{[22]}xi_{[0]} - \xi_{[21]}\xi_{[1]} + \xi_{[2]}\xi_{[11]} = 0,\nn\\
\xi_{[32]}\xi_{[0]} - \xi_{[31]}\xi_{[1]} + \xi_{[3]}\xi_{[11]} = 0,  \nn\\
\xi_{[221]}\xi_{[0]} - \xi_{[211]}\xi_{[1]} + \xi_{[2]}\xi_{[111]} = 0,\nn\\
\ldots
\ee
This indeed happens for one special class \cite{I}: the {\it torus} knots/links
with all $a_{ij}=1$ ($1\leq i\leq m-1,\ 1\leq j\leq n$)
but not in general.

Another possibility is that ${\cal H}^{\cal B}_R$ themselves satisfy the Pl\"ucker relations
as functions of $R$, i.e. that $Z^{\cal B}_R$ is a $\tau$-function w.r.t. times $\bar p_k/k$.
Then, in particular, the Ooguri-Vafa partition function
$\sum_R {\cal H}^{\cal K}_R\{q|A\}S_R\{\bar p_k\}$ would be a KP $\tau$-function in these time variables.
Again, this is not true in general, but an example is known when this
is true (besides the trivial example of the unknot): in the limit $q\rightarrow 1$, when the HOMFLY polynomials turn into
the "special" polynomials
\be
{\mathfrak{S}}_R^{\cal K}(A) = \lim_{q=1} \frac{H_R^{\cal K}(q|A)}{S_R^*(q|A)}
\ee
which have a very simple dependence on $R$:
\be
{\mathfrak{S}}_R^{\cal K}(A) = \Big({\mathfrak{S}}_{_\Box}^{\cal K}(A)\Big)^{|R|}
\ee

The full HOMFLY polynomial is restored by action of a
combination of the cut-and-join operators \cite{spepo}
\be
\sum_R
H_R^{\cal K}(A) S_R\{\bar p_k\}
=\exp \left( \sum_Q (q-q^{-1})^{c_Q}{\spepo_{Q}^{\cal K}(A)\over \left(\spepo_{_\Box}^{\cal K}(A)\right)^{|Q|}} \hat W_Q\{\bar p_k\}\right)
\exp\left(\sum_k \frac{1}{k}\Big(\spepo_{\Box}^{\cal K}(A)\Big)^k p_k^*\bar p_k\right)
\label{spepoexpan}
\ee
with
\be
 \spepo_{_\Box}^{\cal K}(A)\equiv {\mathfrak{S}}_{_\Box}^{\cal K}(A)
 \ee
and resembles very much the Hurwitz partition function \cite{Hur,Hur2},
which is also in general not a KP/Toda $\tau$-function in the variables
$\{p,\bar p\}$.
In the Hurwitz case the way is known to cure this problem \cite{Hurint}
(by switching to times associated with a special basis in the space
of cut-and-join operators), perhaps, this analogy can show a way to proceed
in the case of HOMFLY polynomials as well.
The integer coefficients of the special polynomials $\sigma_{Q}^{\cal K}(A)$
are related to the Ooguri-Vafa numbers \cite{OV,III}, but in a somewhat complicated way
for $R\ne\Box$.

\section{$\hat{\cal A}$-polynomials}

The fourth, currently most interesting type of relations are difference
equations in $R$ \cite{Apol}, so far found empirically mostly for symmetric representations $R=[p]$ \cite{FGS1,IMMMfe,FGS2}
or for the small groups with $N=2,3$ \cite{Gar,Gar3}.

\bigskip

The usual strategy to derive these equations \cite{Gar}
(at least at $N=2$, when $R=[r-1]$ is given by the single positive integer)
is to

$\bullet$ convert the Jones polynomial (with $A=q^2$)
into a form of a proper $q$-hypergeometric series \cite{Noumi,WZ}
\be
J_r(q)=\sum_{{\bf k}\in {\mathopen{Z}}^r}\prod_{i,j}{(B^+_i,q^2)_{b^+(r,{\bf k})}
\over (B^-_i,q^2)_{b^-(r,{\bf k})}}q^{C(r,{\bf k})}{\bf Y}^{\bf k}
\ee
where $b^{\pm}(r,{\bf k})$ and
$C(r,{\bf k})$ are linear forms and a quadratic form accordingly, which, as well as
$B^{\pm}_i$ and the $r$-vector ${\bf Y}$, are
just parameters which describe the concrete knot and the $q$-Pochhammer symbol
\be
(B,q)_r=\prod_{i=0}^{r-1}(1-Bq^i)
\ee

$\bullet$  write down a difference equation, which always exists
for the proper $q$-hypergeometric series.

\bigskip

\noindent
For example, in the simplest case of the trefoil ($3_1$ knot) \cite{Hab}
\be
J_{r}^{3_1}(q)=q^{2-2r}\sum_{k=0}^\infty q^{-2kr}\Big(q^{2-2r},q^2\Big)_k
\ee
the equation is \cite{Gar,3dAGT}:
\be\label{A31}
J_{r}^{3_1}(q) -  U_r(q) J_{r-1}^{3_1}(q) = V_r(q), \nn \\
U_r(q) = -q^{4-6r}{1-q^{2r-2}\over 1-q^{2r}}, \ \ \ \ V_r(q) =q^{4(1-r)}{1-q^{4r-2}\over 1-q^{2r}}
\ee
For arbitrary torus knot $[m,n]$ (trefoil is either $[2,3]$ or $[3,2]$),
the equation is of the second order, however, for the whole series $T[2,2s+1]$ \cite{Hik1}
\be
J_{r}^{T[2,2s+1]}(q)=q^{2s(1-r)}\sum_{k_m\ge\ldots\ge k_1\ge 0}
q^{-2k_mr}\Big(q^{2-2r},q^2\Big)_{k_m}\times\left(\prod_{i=1}^{s-1}q^{2k_i(k_i+1-2r)}
{\Big(q^2,q^2\Big)_{k_{i+1}}
\over \Big(q^2,q^2\Big)_{k_{i}}\Big(q^2,q^2\Big)_{k_{i+1}-k_i}}
\right)
\ee
it reduces to
the first order and looks similar to (\ref{A31}) with \cite{Hik1}
\be
U_r(q) = -q^{2(s+1)-2(2s+1)r}{1-q^{2r-2}\over 1-q^{2r}}, \ \ \ \
V_r(q) =q^{2(s+1)(1-r)}{1-q^{4r-2}\over 1-q^{2r}}
\ee

For non-torus knots the formulas get longer, and it is convenient to
write them in terms of the operator $\hat M$: $\hat Mf(q,r)=q^{-2r}f(q,r)$ and
the shift operator $\hat L$:
$\hat Lf(q,r)=f(q,r+1)$, $\hat L\hat M=q^2\hat M\hat L$.
Then, for knot $4_1$
\be
J_{r}^{4_1}(q)=\sum_{k=0}^\infty (-1)^kq^{-k(k+1)}\Big(q^{2-2r},q^2\Big)_k
\Big(q^{2+2r},q^2\Big)_k
\ee
the equation is \cite{Gar,3dAGT}:
\be\label{differeq}
\hat{\cal A}^{4_1}(\hat L,\hat M) J^{4_1}_r(q) = {\cal B}(\hat M),
\ee
$$
\hat{\cal A} = q^4\hat M^2(1-\hat M)(1-q^6\hat M^2)-(q^2\hat M+1)(1-q^2\hat M-q^2\hat M^2-q^6
\hat M^2-
q^6\hat M^3+q^8\hat M^4)(1-q^2\hat M)^2\hat L+
$$
$$
+q^4\hat M^2(1-q^2\hat M^2)(1-q^4\hat M)\hat L^2, \ \ \ \ \
\hat{\cal B} = q^2\hat M(1-q^6
\hat M^2)(1-q^2\hat M^2)(1+q^2\hat M)
$$
In the limit of $q\to 1$, $\hat {\cal A}(\hat L,\hat M) $ can be considered as
an ordinary $c$-number function ${\cal A}(l,m)$
of $\hat L\to l$ and $\hat M\to m$ and coincides (this is called the
AJ-conjecture \cite{Apol}) with the ${\cal A}$-polynomial
\cite{CCGLS} (this is why $\hat{\cal A}^{{\cal K}}(\hat L,\hat M)$ is called the non-commutative
${\cal A}$-polynomial),
which defines a "spectral curve", associated with the knot ${\cal K}$:
\be
\Sigma^{\cal K}: \ \ \ \ {\cal A}^{\cal K}(l,m) = 0
\label{specu}
\ee
which is naturally equipped with the Seiberg-Witten differential
\be
dS^{\cal K} = \log l\ d\log m
\label{SWdiff}
\ee
According to the standard dictionary of the Seiberg-Witten theory \cite{SWint,SW5d,GGM2},
this form of the SW differential
is a clear sign of the relation to $5d$ YM theories.

\section{The large-$|R|$ limit}

Given a difference equation (\ref{differeq}), a natural question to
ask is about various kinds of the loop (quasiclassical) expansions
for its solutions. Loop expansion arises when one introduces the 't
Hooft coupling constant $u$ by putting $q = e^{\hbar} = e^{g^2} =
e^{u/|R|}$ and $|R| = \frac{u}{h} = \frac{u}{g^2}$\ .

\subsection{Special polynomial expansion}

To begin with one can look at the anzatz, provided by the
special polynomial expansion (\ref{spepoexpan}). It would imply that the
solution behaves like
\be
{\cal H}_R^{{\cal K}}\sim \spepo^{{\cal K}}_\Box(A)^{|R|} = \exp
\Big(|R|\cdot \log\sigma^{{\cal K}}_\Box(A)\Big), \label{asympt}
\ee
i.e. grows
exponentially with $|R|$, in accordance with the volume conjecture
\cite{VC}. However, things are not so simple and strongly depends on the range of values of $u$.

First of all, corrections to this formula could also contribute to
the exponential growth. Indeed, the next correction is \cite{spepo}
\be
\exp\left(\epsilon\sum_R
\frac{\varkappa_R\spepo^{{\cal K}}_{[2]}(A)}{\spepo_\Box(A)}\right)
\ee
where
$\varkappa_R \sim \frac{|R|^2}{2} = \frac{r^2}{2}$ at large $|R|=r$ and
$\epsilon = q-q^{-1}$. Since the latter is equal to $2\sinh \hbar = 2\sinh g^2 \sim
\frac{u}{|R|}$, the exponent in this formula is proportional to
$\epsilon \varkappa_R \sim |R|$ at large $|R|$, so that one should
add $\frac{u\spepo^{{\cal K}}_{[2]}(A)}{\Big(\sigma^{{\cal K}}_\Box(A)\Big)^2}$ to $\log\
\spepo_\Box(A)$ in (\ref{asympt}). The next correction will be
proportional to $\frac{u^2\spepo^{{\cal K}}_{[3]}(A)}{\Big(\spepo^{{\cal K}}_\Box(A)\Big)^3}$ and so
on: there is an infinite series in powers of $u$ of contributions to
the leading ($\sim |R|$) "volume" behavior of the exponent.

Second, for $N$ fixed, i.e. for $A\to 1$ (for example, for the Jones polynomial with
$A=q^2$), this volume contribution is in fact vanishing term by term,
because $\spepo^{{\cal K}}_\Box(A=1) = 1$ and all other $\spepo^{{\cal K}}_Q(A=1)=0$
\cite{spepo}. In fact this is in accordance with the well known fact
that for small enough $u$ there is no exponential growth in
$J_r(q=e^{u/r})$. Instead (this formula was realized for the figure eight in
\cite{JA1} and for generic knots in \cite{JA2} basing on the
Melvin-Morton-Rozansky conjecture \cite{MMR})
\be
J_r(q=e^{u/r}) =
\exp\left(\sum_{k\geq 0} \left(\frac{u}{r}\right)^{2k}f_k(u)\right)
= \frac{1}{{\rm Alexander}(q=e^u)} + \sum_{k\geq 1} r^{-k}
\frac{w_k(q=e^u)} {\Big({\rm Alexander}(q=e^u)\Big)^k}
\ee
with some
polynomials $w_k$. In particular,
\be
f_0(u) = - \log\Big({\rm
Alexander}(q=e^u)\Big)
\ee
and this is in a nice accordance with the
special polynomial expansion \cite{spepo}.

Only when $e^u$ exceeds the smallest
root of the Alexander polynomial, another solution appears, with
$f_{-1}(u)\neq 0$.

At the same time, if one chooses $u=2\pi i$, there is the
exponential behaviour (\ref{asympt}) \cite{VC1} and the resulting coefficient
in front of $|R|$, for the Jones polynomial $A=q^2$, is equal to the
hyperbolic volume of the knot \cite{VC} ({\it volume conjecture}).

\subsection{Loop expansion}

It is described by the standard loop expansion:
\be
J_r(q=e^{g^2}) = \exp \left(\sum_{k\geq -1} g^{2k}f_k(v)\right)
\ee
where $v=u-2\pi i$, i.e. $v=0$ is the volume conjecture point.
The shift operations are
\be
J_{r\pm 1}(q=e^{g^2}) = \exp
\left(\sum_{k\geq -1} g^{2k}f_k(v\pm g^2)\right) =
\exp\left(\sum_{\stackrel{j=0}{k\geq -1}} (-)^j g^{2k+2j}
\partial^j_v f_k(v)\right)
\ee
Therefore, the difference equation
(\ref{differeq}), if expanded in powers of $g^2$, turns into an
infinite system of recurrent relations for infinite set of functions
$f_k(v)$. Remarkably this system can be related \cite{DFM} to a
similar system of equations for multidensities
$\rho_{p,k}(v_1,\ldots,v_k)$, which constitutes the essence of the
AMM/EO topological recursion \cite{AMMEO}. More exactly, the knot polynomial is
associated with $\Big<\log\det (M-v)\Big>$, where the brackets
denote averaging over the "matrix" $M$ in the would-be matrix model
underlying the recursion.

Recursion itself means that, given $f_{-1}(v)$, usually described
in terms of the spectral Riemann surface (\ref{specu}) with a
Seiberg-Witten differential (\ref{SWdiff}), one can recursively
reconstruct all the functions $f_k(v)$.

\section{$3d/5d$ AGT relation}

\subsection{${\cal A}$-polynomials, state integral model and Baxter equations}

As soon as $f_{-1}\neq 0$ appears, one can neglect the r.h.s. in
equation (\ref{differeq}) and substitute it by the homogeneous ${\cal
A}$-polynomial equation. This one has a simpler solution, in the
form of integrals of the double-sine ratios instead of sums of the
$q$-Pochhammer ratios \cite{Hik}. This solution
is known as partition function of the state integral model \cite{Zag}.

According to \cite{3dAGT} one can interpret this homogeneous
difference equation as a Baxter equation for some relativistic
integrable system (such systems are associated through
Seiberg-Witten theory \cite{SW,SWint,SW5d,GGM2} to $5d$  SYM theories, what
explains the possible name for such a relation).

In fact, this should not come as a surprise due to the well-known
connection between $2d$ CFT and Chern-Simons theory. As soon as the
former one is related via the AGT correspondence \cite{AGT,AGTmore,AGTrev} with the $4d$ SW
systems, one would naturally expect a similar relation in the
dimensions increased by unit: the one between Chern-Simons theory
and $5d$ SW theory\footnote{Within a different context, a $5d/3d$ duality emerged in
\cite{CHZ}.}. On the other hand, the Baxter equation is an
evident object to look at, since it encodes changing traces of the
group element with representation (fusion), and there is a
counterpart of this on the knot theory side: the difference ${\cal
A}$-polynomial equation describes exactly changing the traces (of
the braid group element) with representation. The technical
similarity of these two things is however appealing.

To be more concrete, one can notice that eq.(\ref{differeq}) has the
form of the Baxter equation for the relativistic Toda chain, which,
in accordance with \cite{SW5d}, describes the $5d$ SW system.
Instead, one could definitely start from the classical limit of the
Baxter equation (relativistic Toda spectral curve) noticing that it
coincides with the "classical" ${\cal A}$-polynomial, and then apply
the topological recursion with the $5d$ SW differential
$dS_{SW}$ (\ref{SWdiff}). This was done in \cite{DFM} and the result reproduced
the expected asymptotics of the Jones polynomial (see also \cite{GS}
for a more generic context).

This is depicted at the scheme:

\be
\begin{array}{ccccc}
\hat{\cal A} H = 0 && \longrightarrow && 5d\ {\rm Baxter}(Q) = 0 \\
&& && d{\cal S}=d\log\Psi \\
\\
\uparrow &&\boxed{H=Q}&& \downarrow \ \hbar=0 \\
\\
H &&  {\rm AMM/EO\ topological\ recursion}  && \Sigma:\ {\cal A}=0\\
&&\longleftarrow&& dS_{SW}
\end{array}
\ee

This scheme implies that, at the quantum level, there is a
relativistic integrable system/$5d$ SW system described by the
Baxter equation for the Baxter operator $Q(\mu)$. The Fourier
transform of it, $\Psi(x)$ satisfies a relativistic Schr\"odinger
equation, and one can construct the $5d$ Nekrasov functions (in the
Nekrasov-Shatashvili limit \cite{NS}) as the Bohr-Sommerfeld
integrals of this equation \cite{BS}, which are A- and B-periods of
the differential $d{\cal S}=d\log\Psi (x)$. On the other hand, the
same $Q(\mu)$ is the partition function of the state integral model,
since this latter satisfies the same difference equation.

At the classical level, the system is described by the Riemann
surface (the spectral curve of the classical integrable system) of
the $5d$ SW system with the differential given by the classical
limit of $d{\cal S}$: $d{\cal S}\stackrel{\hbar\to
0}{\longrightarrow} dS_{SW}$. This Riemann surface coincides with
the classical ${\cal A}$-polynomial (which describes the $SL(2)$ representation
of the knot complement to $S^3$, as viewed from the boundary \cite{CCGLS})
and $dS_{SW}$ (\ref{SWdiff})
allows one to apply the topological recursion and to restore the
partition function of the state integral model

\subsection{Universal group element}

Let us now briefly describe the origin of the Baxter equation as the fusion relation of
the universal group element. First
of all, one has to construct the group element given over the
non-commutative ring, the algebra of functions on the quantum group
(see a review in \cite{Mir}). That is, one constructs such an
element $g\in U_q({\cal G})\otimes A(\G)$ of the tensor product of
the Universal Enveloping Algebra (UEA) $U_q({\cal G})$ and its dual
algebra of functions $A(\G)$ that \be\label{Tcoprod}
\Delta_U(g)=g\otimes_{U} g \in A(\G)\otimes U_q({\cal G})\otimes
U_q({\cal G}) \ee To construct this element \cite{FRT,Mir},
we fix some basis $T^{(\alpha)}$ in $U_q({\cal G})$. There exists a
non-degenerated pairing between $U_q({\cal G})$ and $A(\G)$, which
we denote $<...>$. We also fix the basis $X^{(\beta)}$ in $A(\G)$
orthogonal to $T^{(\alpha)}$ w.r.t. this pairing. Then, the sum
\be\label{grouel} \hbox{{\bf
T}}\equiv\sum_{\alpha}X^{(\alpha)}\otimes T^{(\alpha)}\in A(\G)
\otimes U_q({\cal G}) \ee is exactly the group element we are
looking for. It is called the universal {\bf T}-matrix (as it is
intertwined by the universal ${\cal R}$-matrix) or the universal
group element.

In order to prove that (\ref{grouel}) satisfies formula
(\ref{Tcoprod}) one should note that the matrices
$M^{\alpha\beta}_{\gamma}$ and $D^{\alpha}_{\beta\gamma}$ giving
respectively the multiplication and co-multiplication in $U_q({\cal
G})$ \be T^{(\alpha)}\cdot T^{(\beta)}\equiv
M^{\alpha\beta}_{\gamma}T^{(\gamma)},\ \ \Delta(T^{(\alpha)})\equiv
D^{\alpha}_{\beta\gamma}T^{(\beta)}\otimes T^{(\gamma)} \ee give
rise to, inversely, co-multiplication and multiplication in the dual
algebra $A(\G)$: \be\label{DM}
D^{\alpha}_{\beta\gamma}=\left<\Delta(T^{(\alpha)}),X^{(\beta)}\otimes
X^{(\gamma)}\right>\equiv \left<T^{(\alpha)},X^{(\beta)}\cdot
X^{(\gamma)}\right>\\
M^{\alpha\beta}_{\gamma}=\left<T^{(\alpha)}T^{(\beta)},X^{(\gamma)}\right>=
\left<T^{(\alpha)}\otimes T^{(\beta)},\Delta(X^{(\gamma)})\right>\nn
\ee Then, \be \Delta_U(\hbox{{\bf
T}})=\sum_{\alpha}X^{(\alpha)}\otimes \Delta_U(T^{(\alpha)})=
\sum_{\alpha,\beta,\gamma}D^{\alpha}_{\beta\gamma}X^{(\alpha)}\otimes
T^{(\beta)}\otimes
T^{(\gamma)}=\sum_{\beta,\gamma}X^{(\beta)}X^{(\gamma)} \otimes
T^{(\beta)}\otimes T^{(\gamma)}=\hbox{{\bf T}}\otimes_U\hbox{{\bf
T}} \ee This is the first defining property of the universal {\bf
T}-operator, which coincides with the classical one. The second
property that allows one to consider {\bf T} as an element of the
"true" group is the group multiplication law $g\cdot g'=g''$ given
by the map: \be\label{grouplaw} g\cdot g'\equiv \hbox{{\bf
T}}\otimes_A \hbox{{\bf T}}\in A(\G)\otimes A(\G)\otimes U_q({\cal
G}) \longrightarrow g''\in A(\G)\otimes U_q({\cal G}) \ee This map
is canonically  given by the co-multiplication and is again the
universal {\bf T}-operator: \be \hbox{{\bf T}}\otimes_A\hbox{{\bf
T}}=\sum_{\alpha,\beta}X^{(\alpha)}\otimes X^{(\beta)}\otimes
T^{(\alpha)}T^{(\beta)}=\sum_{\alpha,\beta,\gamma}
M^{\gamma}_{\alpha,\beta}X^{(\alpha)}\otimes X^{(\beta)}\otimes
T^{(\gamma)}= \sum_{\alpha}\Delta(X^{(\alpha)})\otimes T^{(\alpha)}
\ee i.e.
$$
g\equiv\hbox{{\bf T}}(X,T),\ \ g'\equiv \hbox{{\bf T}}(X',T),\ \
g''\equiv\hbox{{\bf T}}(X'',T)
$$
\be \ \ X\equiv\{X^{(\alpha)}\otimes I\}\in A(\G)\otimes I, \ \
X'\equiv\{I\otimes X^{(\alpha)}\}\in I\otimes A(\G)\ \
X''\equiv\{\Delta(X^{(\alpha)})\}\in A(\G)\otimes A(\G) \ee

\subsection{Baxter equation}

The most essential property of the group elements is that they are
intertwined by the universal $R$-matrix \cite{URmat}, i.e.
\be
{\cal
R}\ \hbox{{\bf T}}\otimes_U\hbox{{\bf T}}'=\hbox{{\bf
T}}'\otimes_U\hbox{{\bf T}}\ {\cal R}
\ee
This means that their
traces taken at any representations of the UEA, $\T_V$ commute:
\be\label{Tcom}
\left[ \T_V,\T_{V'}\right]=0
\ee

The equation that connects these traces in different representation
is called fusion and its specific limit is the Baxter equation. As a
simple instance, let us consider the case of quantum affine algebra
$\widehat {SL(2)}_q$ and the evaluation representation of this
latter $V_j(\lambda)$ labeled by the spin $j$ and evaluation
parameter $\lambda$ (see, e.g., \cite{Jimbo} for a
review)\footnote{Remind that these representations can be treated as
either finite dimensional but reducible, or infinite dimensional
irreducible representations \cite{Jimbo}. We prefer here the latter
point of view.}. The product of two such representations
$(j,\lambda)$ and $(k,\mu)$ is an irreducible representation unless
\be
{\lambda\over\mu}=q^{j+k-p+1}
\ee
where $p=1,...,min(2j,2k)$.
In this latter case, the product is partly reducible:
\be
0\longrightarrow V_{j-p/2}(q^{p/2}\lambda)\otimes
V_{k-p/2}(q^{-p/2}\mu)\longrightarrow V_j(\lambda)\otimes
V_k(\mu)\longrightarrow\nn\\ \longrightarrow V_{j+k-p/2+1/2}(q^{p/2-k-1/2}\lambda)\otimes
V_{p/2-1/2}(q^{k+1/2-p/2}\mu) \longrightarrow 0
\ee
In terms of traces
of the group element, this (fusion) relation can be written in the
form
\be\label{fusion}
\T_j(\lambda)\T_k(\mu)=\T_{j-p/2}(q^{p/2}\lambda)\T_{k-p/2}(q^{-p/2}\mu)+
\T_{j+k-p/2+1/2}(q^{p/2-k-1/2}\lambda)\T_{p/2-1/2}(q^{k+1/2-p/2}\mu)
\ee
Choosing $k=1/2$, $p=1$ and $\lambda=q^{j+1/2}\mu$, one
immediately obtains
\be\label{fusion2}
\T_j(q^{j+1/2}\mu)\T_{1/2}(\mu)=\T_{j-1/2}(q^{j+1}\mu)+\T_{j+1/2}(q^{j}\mu)
\ee
It turns out \cite{AF} that (upon the proper normalization of
$\T_j(\lambda)$) there is a finite limit
\be
\lim_{j\to
\infty}\T_j(q^{j+1/2}\mu)=Q(\mu)
\ee
and (\ref{fusion2}) turns into
the standard Baxter equation
\be\label{Bax1}
\T_{1/2}(\mu)Q(\mu)=Q(q\mu)+Q(q^{-1}\mu)
\ee
Here $\T_{1/2}(\mu)$
is a Laurent polynomial of degree 1 both in $\mu$ and $\mu^{-1}$.
One can generate Laurent polynomials of higher degrees considering
in (\ref{fusion2}) traces of products of the group elements,
(\ref{grouplaw}), the product of $n$ group elements giving rise to
degree $n$. Thus, generically one would get instead of (\ref{Bax1})
\be\label{Bax1n}
P_n[\mu,\mu^{-1}]Q(\mu)=Q(q\mu)+Q(q^{-1}\mu)
\ee
which coincides with the Baxter equation for the $n$-particle
relativistic Toda chain, which corresponds at the SW level to the
$5d$ pure gauge theory. The transition to four dimensions and the
standard Toda system is done by replacing the affine algebra with
the Yangian. This leads to the same equation with $P_n$ being a
degree $n$ polynomial of $\mu$.

\subsection{Knots, Baxter equations and $5d$ SW theory}

One can also choose $p=2$ and $k=3/2$. Then,
\be\label{B2}
\boxed{Q(\mu)\T_{3/2}(\mu)=\T_{1/2}(q^{-1}\mu)Q(q^2\mu)+\T_{1/2}(q\mu)Q(q^{-2}\mu)}
\ee
This equation is another version of the Baxter equation, it gives rise to the same
spectral curve in the classical limit.

Let us consider, for instance,
the first non-trivial example of the figure eight knot $4_1$ which ${\cal A}$-polynomial
equation is the homogeneous part of (\ref{differeq}), i.e.
\be\label{B41}
\hat{\cal A}^{4_1}(\hat L,\hat M) {\mathfrak{J}}^{4_1}(M)=0
\ee
We use the basis where $\hat M$ acts on ${\mathfrak{J}}$ diagonally so that (\ref{B41})
reduces to
\be\label{new}
(1-q^6\mu^2)Q(q^{-2}\mu)+(1-q^2\mu^2)Q(q^{2}\mu)=P_2[\mu,\mu^{-1}]Q(\mu)\ ,\nn\\
P_2[\mu,\mu^{-1}]=\Big(q^4\mu^2+{1\over q^4\mu^2}
\Big)-\Big(q^2\mu+{1\over q^2\mu}\Big)-\Big(q^2+{1\over q^2}\Big)
\ee
where $Q(\mu)=(1-\mu){\mathfrak{J}}^{4_1}(\mu)$ with identification $M=\mu$. (\ref{new}) is
nothing but (\ref{B2}) with
\be\label{83}
\boxed{
\T_{1/2}(\mu)=q^2\mu-{1\over q^2\mu}\ ,\ \ \ \ \ \ \ \ \ \ \ \ \
\T_{3/2}(\mu)=\left(q^2\mu-{1\over q^2\mu}\right)P_2[\mu,\mu^{-1}]}
\ee
The coefficients of $\T_{1/2}(\mu)$ and $\T_{3/2}(\mu)$ can be
related with concrete values of the energies (integrals of motion) in the periodic 2-particle
relativistic Toda chain, and the corresponding SW curve ($5d$ pure gauge theory) is
\be\label{SWcurve}
l+{1\over l}=\Big(\mu^2+{1\over\mu^2}\Big)
-\Big(\mu+{1\over \mu}\Big)-2
\ee
In contrast with the generic SW curve, where there is an arbitrary degree two Laurent
polynomial at the r.h.s. of (\ref{SWcurve}), a peculiar polynomial with all coefficients fixed
corresponds to the figure eight knot. Similarly, (\ref{83}) specifies very concrete polynomials, while
in the generic integrable system these are arbitrary ones (of fixed degree).

Note that choosing $k=1$, $p=1$ and $\lambda=q^{j+1}\mu$ (or $p=2$
and $\lambda=q^{j}\mu$), one obtains from (\ref{fusion}) a third
order difference equation (or its counterpart with $q\to 1/q$)
\be
\T_{1}(q^{1/2}\mu)Q(\mu)=Q(q^2\mu)+\T_{1/2}(q\mu)Q(q^{-1}\mu)
\ee
The
procedure can be continued further to generate higher order
difference equations which can be compared with the ${\cal
A}$-polynomial equations for other, more complicated knots. In
accordance with \cite{Gar}, such an equation exists for any knot.
The coefficients of the polynomial $P_n[\mu,\mu^{-1}]$ are fixed by
choosing the representation of the algebra of functions on the
quantum group. Indeed, due to (\ref{Tcom}) we know that all the
coefficients of this polynomial are commuting. They are nothing but
the Casimir functions that determine the representation of $A(\G)$.

In integrable systems these coefficients are just integrals of
motion, which can have arbitrary values. On the contrary, in knot
theory their values are fixed to be very concrete numbers, i.e. the
representations of $A(\G)$ in knot theory are fixed. It still
remains unclear what is the condition that fixes them in group
theory.

Now let us stress out that all these Baxter equations or quantum
${\cal A}$-polynomials correspond in the classical limit to $5d$ SW
theories, generically of the quiver type \cite{Wit,Gaiotto}. On the
integrable side, these theories are described by various
degenerations of the $sl(p)$ $XXZ$ spin chain \cite{GGM2} (or
corresponding Gaudin type systems, see \cite{Zotov}). Their
quantization is immediate, either as integrable systems which goes
exactly through the Baxter equation \cite{BS}, or as ${\cal
A}$-polynomials \cite{DFM,GS}.

Even more general Baxter equations are described by the generic
$sl(p)$ $XXZ$ spin chain, in the classical limit giving rise to the
$5d$ SW curves that are plane complex curves given by sets of
arbitrary Laurent polynomials $P_{n_i}[\mu,\mu^{-1}]$ \cite{GGM2}
\be
\sum_{i=1}^pw^iP_{n_i}[\mu,\mu^{-1}]=0
\ee
However, obtaining these Baxter equations directly from the representation theory as in
the previous subsection is not that immediate in this case, and the whole construction
becomes less elegant and much more involved \cite{DM}.

Note that one can similarly consider the Baxter equation describing
$\widehat {SL(N)}_q$. In terms of knot theory, it should describe
the difference equations for the HOMFLY polynomials with
specialization $A=q^N$.

\subsection{State integral model and open relativistic Toda chain}

Let us note that solutions to the Baxter equations are usually
quite involved, and are unknown in a generic case. On the contrary, the partition function of
the state integral model, which solves the ${\cal A}$-polynomial (i.e. Baxter)
equation is often known. The reason is that with peculiar values of the integral of motions, i.e. with
the specific form of traces $\T_{j}$ corresponding to knots, the equation is explicitly solved. The answer is
presented as a multiple integral of product of the quantum dilogarithms \cite{KLS}. This answer is in fact
associated with the wave function of the open relativistic Toda chain (which, within the SW context, corresponds
to the perturbative limit of the gauge theory). More exactly, this wave function solves the Schr\"odinger
equation given by the Hamiltonian of the system {\it dual} \cite{dual,dualc,dualq}
to the relativistic Toda chain. In other words,
this is the (difference) equation that describes the dependence of the wave function on the energies.

It happens, since the Baxter equation, which is a difference equation for the separated variables in the
periodic chain, is simultaneously (at peculiar values of energies) a Sch\"odinger equation for the (dual) open chain
(this is possible definitely only due to the fact that chain is relativistic and, hence, both the Schr\"odinger
and the Baxter equations are difference ones).

Let us see how it works in already discussed simplest non-trivial example of the figure eight knot.
Remind that the Hamiltonian of the $k$-particle relativistic Toda
chain is given by \cite{Rui}
\be
h\{x_i;p_i\}=\sum_{i=1}^k\left[e^{\omega_1 p_i}+q^{-1}g^{2\omega_1}e^{{2\pi\over\omega_2}(x_i-x_{i+1})
+{1-\epsilon\over 2}\omega_1 p_i+{1+\epsilon\over 2}\omega_1 p_{i+1}}\right]
\ee
with the boundary condition $x_{k+1}=x_1$ in the periodic case and $x_{k+1}=\infty$ in the open case. Here
$q=e^{\pi i\omega_1/\omega_2}$ is the relativistic parameter ($q\to 1$ as the speed of light goes to infinity),
$\epsilon=0,\pm 1$ parameterizes different relativistic Toda systems
and $g$ is a coupling constant, which can be removed by redefinitions of coordinates $x_i$ only in the open case.
In the 2-particle open case, the Schr\"odinger equation for $\epsilon=0$
can be reduced (upon neglecting the $U(1)$-factor) to the equation
\be
\psi_\gamma (q^{-1}\mu)+\psi_\gamma (q\mu)+g^{2\omega_1}\mu\psi_\gamma (\mu)=\left(e^{\pi\gamma}+e^{-\pi\gamma}
\right)\psi_\gamma (\mu)
\ee
where $\gamma$ parameterizes the energy and we fixed $\omega_2=1$, $\omega_1=\hbar/\pi i$ and denoted $\mu=\exp(2\pi x)$,
$x=x_1-x_2$.
A solution to this eigenvalue problem is of the form \cite{KLS}
\be\label{todasol}
\psi_\gamma (\mu)\propto \int_C dt\ {{\cal S}\Big(-it-ix+1+{\hbar\over\pi i}-
{\hbar\over 2\pi^2}\log g\Big|{\hbar\over \pi i},1\Big)
\over {\cal S}\Big(it+i{\gamma\over 2}-{\hbar\over 2\pi^2}\log g\Big|{\hbar\over \pi i},1\Big)}\
\displaystyle{e^{-{\pi ^2\gamma (2t+x)\over\hbar}}}
\ee
where $S(x|\hbar/\pi i,1)$ \cite{KLS} is related with the quantum dilogarithm \cite{Zag} by the formula
\be
\Phi_\hbar(z)={\cal S}\Big({z\over 2\pi i}+{\pi i+\hbar\over 2\pi i}\Big|{\hbar\over \pi i},1\Big)
\ee
and the contour $C$ is defined in \cite[(2.26)]{KLS}.
Now, if one puts $\mu g^{2\omega_1}=1$,
integral (\ref{todasol}) reduces to
(${p\over 2\pi}=t-{\hbar\over 2\pi^2 i}\log g$)
\be
\psi_\gamma (g^{-2\omega_1})\propto\int_Cdp\ {\Phi_\hbar(p+i\pi+\hbar)\over\Phi_\hbar (-p-\pi\gamma-i\pi-\hbar)}
e^{-{\pi\gamma\over\hbar}(p+\pi\gamma/2)-\pi\gamma/2}
\ee
which is exactly the partition function of the state model for the figure eight knot as a function of
$u=\pi\gamma/2$ \cite{Zag}
\be
\boxed{{\mathfrak{J}}^{4_1}(e^u)\propto\psi_{2u/\pi} (g^{-2\omega_1})}
\ee
Since the dependence on $\gamma$ is regulated by the dual Toda Hamiltonian, we come to the claim above.

In the case of more complicated knots one has instead of (\ref{todasol})
a multiple integral of the product of the quantum dilogarithms \cite{Zag}. This integral could be associated with multi-particle
relativistic open Toda wave functions \cite{KLS} and their further generalizations.

\section{Conclusion}

To conclude, in development of the program outlined in \cite{3dAGT,I},
we presented a brief review of the new and very interesting research field:
the study of interrelations between various HOMFLY polynomials,
probably reflecting the powerful underlying integrable structure,
which still remains to be fully revealed.
The relations involve different knots/links and different representations,
they are rather diverse and form a somewhat strange intertwined set,
and partly unexpected parallels are seen with the Hurwitz theory
(i.e. with that of symmetric group characters),
quantum integrable systems and AGT relations.
Even more interesting should be generalizations from the HOMFLY to
{\it super}polynomials \cite{sp}, however, what structures survive this $\beta$-deformation \cite{beta}
remains an open question.

A way to systematical study of these problems is now open by the
powerful HOMFLY calculus, developed in \cite{DMMSS,I,II,III,Ano},
exploiting the simple and universal structure of $SU_q$
$R$-matrices and Racah coefficients,
and allowing one to obtain the parameter-dependent expressions for
knot polynomials.
One now expects a fast and fruitful development of this old and
intriguing field.

\section*{Acknowledgements}

The authors are grateful to S.Kharchev for valuable discussions.

Our work is partly supported by Ministry of Education and Science of
the Russian Federation under contract 14.740.11.0677, by NSh-3349.2012.2,
by RFBR grants 10-01-00536 and
by joint grants 11-02-90453-Ukr, 12-02-91000-ANF, 12-02-92108-Yaf-a,
11-01-92612-Royal Society.

\end{document}